\documentclass[twocolumn,showpacs,prl,preprintnumbers,amsmath,amssymb,superscriptaddress]{revtex4}

\usepackage{graphicx}
\usepackage{dcolumn}
\usepackage{bm}
\usepackage{hyperref}
\usepackage{latexsym}

\begin{document}

\fontfamily{ptm}
\selectfont

\title{Optimal network topologies for local search with congestion}

\author{R. Guimer\`a}

\affiliation{Departament d'Enginyeria Qu\'{\i}mica, Universitat Rovira
i Virgili, 43007 Tarragona, Spain}

\author{A. D\'{\i}az-Guilera}

\affiliation{Departament de F\'{\i}sica Fonamental, Universitat de
Barcelona, 08028 Barcelona, Spain}

\affiliation{Departament d'Enginyeria Qu\'{\i}mica, Universitat Rovira
i Virgili, 43007 Tarragona, Spain}

\author{F. Vega-Redondo}

\affiliation{Departament de Fonaments d'An\`alisi Econ\`omica,
Universitat d'Alacant, 03071 Alacant, Spain}

\affiliation{Departament d'Economia i Empresa, Universitat Pompeu
Fabra, 08005 Barcelona, Spain}

\author{A. Cabrales}

\affiliation{Departament d'Economia i Empresa, Universitat Pompeu
Fabra, 08005 Barcelona, Spain}

\author{A. Arenas} 

\affiliation{Departament d'Enginyeria Inform\`atica i Matem\`atiques,
Universitat Rovira i Virgili, 43007 Tarragona, Spain}

\begin{abstract}
The problem of {\it searchability} in decentralized complex networks
is of great importance in computer science, economy and sociology. We
present a formalism that is able to cope simultaneously with the
problem of search and the congestion effects that arise when parallel
searches are performed, and obtain expressions for the average {\it
search cost}---written in terms of the search algorithm and the
topological properties of the network---both in presence and absence
of congestion. This formalism is used to obtain optimal network
structures for a system using a local search algorithm. It is found
that only two classes of networks can be optimal: star-like
configurations, when the number of parallel searches is small, and
homogeneous-isotropic configurations, when the number of parallel
searches is large.
\end{abstract}


\date{\today}

\maketitle


Recently, the study of topological and dynamical properties of complex
networks has received a lot of interest
\cite{watts98,barabasi99,amaral00}. Part of this interest comes from
the attempt to understand the topology and behavior of computer based
communication networks such as the Internet \cite{faloutsos99} and the
World Wide Web \cite{albert99,huberman99}. However, the study of
communication processes in a wider sense is of interest in other
fields, remarkably the design of organizations
\cite{radner93,bolton94,garicano00}.

One of the general principles that has been discovered in many such
complex networks is the short average distance between nodes
\cite{watts98}. More surprisingly, it has been shown that these short
paths can be found with {\it essentially local} strategies, i.e. with
strategies that do not require precise global information of the
network. Indeed, for social networks, this fact was experimentally
confirmed a long time ago by the famous experiment of Travers and
Milgram \cite{travers69} and theoretical explanations have been given
by Kleinberg \cite{kleinberg00} and, more recently, by Watts {\it
et. al.} \cite{watts02}. These explanations are based on the plausible
assumption that there is a {\it structure} (social, geographical,
etc.) that underlies the complex social network and provides
information that can be exploited heuristically in a search
process. In scale-free communication networks and in some
decentralized peer-to-peer communication networks such as Gnutella or
Freenet, it has been shown \cite{adamic01,tadic01} that the skewness
of the degree distribution and the existence of highly connected hubs
allows the design of algorithms that search quite efficiently even
when the size of the system is large.

Our approach in the present work is complementary to these
efforts. The question we pose is the following: given a search
algorithm that uses purely local information---i.e. knowledge of the
first neighbors in the network---and a fixed set of resources---i.e. a
fixed number of nodes and links---, which is the topology that
optimizes the search process? We consider a general situation where
the network has to tackle several simultaneous (or parallel) search
problems, which in turn rises the important issue of congestion
\cite{jacobson88,arenas01,ohira98,sole01} at overburdened
nodes. Indeed, for a single search problem the optimal network is
clearly a highly polarized star-like structure. This structure is {\it
cheap} to assemble in terms of number of links and efficient in terms
of searchability, since the average cost (number of steps) to find a
given node is always bounded (2 steps), independently of the size of
the system. However, the polarized star-like structure will become
inefficient when many search processes coexist in parallel in the
network, due to the limitation of the central node.

The discovery of optimal structures will be a useful guide to design,
redesign and drive the evolution of communication networks such as
peer-to-peer networks, distributed databases, and organizations.

In this paper we present a formalism that is able to cope with search
and congestion simultaneously, allowing the determination of optimal
topologies. This formalism avoids the problem of simulating the
dynamics of the search-communication process which turns out to be
impracticable, specially close to the congestion point where search
costs (time) diverge. We do not focus on detailed models of any of the
above mentioned communication networks (organizations, computer
networks, etc). Rather, we study a general scenario applicable to {\it
any} communication process. First we calculate the average number of
steps (search cost) needed to find a certain node in the network given
the search algorithm and the topology of the network. The calculation
is exact if the search algorithm is Markovian. Next, congestion is
introduced assuming that the network is formed by nodes that behave
like queues, meaning that are able to deliver a finite number of
packets at each time step \cite{allen90,ohira98,arenas01}. In this
context, we are able (i) to calculate explicitly the point at which
the arrival rate of packets leads to network collapse, in the sense
that the average time needed to perform a search becomes unbounded,
and (ii) to determine, below the point of collapse, how the average
search time depends on the rate at which search process are
started. In both cases, the relevant quantities are expressed in terms
of the topology of the network and the search algorithm. Finally we
obtain optimal structures by performing exhaustive generalized
simulated annealing \cite{tsallis94,penna95} in the space of the
networks with fixed size and connectivity. We find that when the
number of parallel searches is small, the star-like configuration
turns out to be optimal as expected, while for a large number of
parallel searches, a very decentralized and uniform network is
best. Surprisingly, no other structures apart from these
extremely-centralized and extremely-decentralized networks are found
to be optimal.

\bigskip

First, we consider the average cost to find a given node in an
arbitrary communication network when there is no
congestion. Specifically, we focus on a single {\it information
packet} at node $i$ whose destination is node $k$, i.e. a packet
searching for $k$. The probability for the packet to go from $i$ to a
new node $j$ in its next movement is $p_{ij}^k$. In particular,
$p_{kj}^k=0$ $\forall j$ so that the packet is {\it removed} as soon
as it arrives to its destination. The precise form of $p_{ij}^k$ will
depend on the search algorithm. In particular, when the search is
Markovian, $p_{ij}^k$ does not depend on previous positions of the
packet. In this case, the probability of going from $i$ to $j$ in $n$
steps is given by
\begin{equation}
P_{ij}^k(n)=\sum_{l_1,l_2,\dots,l_{n-1}}p_{il_1}^k p_{l_1l_2}^k\cdots p_{l_{n-1}j}^k.
\end{equation}
Thus defining the matrices $\mathbf{p}^k$ and $\mathbf{P}^k(n)$ we have
\begin{equation}
\mathbf{P}^k (n)=\left(\mathbf{p}^k\right)^n.
\end{equation}

We next define the {\it effective} distance matrices
\begin{equation}
\mathbf{d}^k=\sum_{n=0}^\infty n \mathbf{P}^k(n)=\sum_{n=0}^\infty n
\left(\mathbf{p}^k\right)^n=\left[(\mathbf{I}-\mathbf{p}^k)^{-1}\right]^2\mathbf{p}^k,
\end{equation}
whose elements $d_{ij}^k$ are the average number of steps needed to go
from $i$ to $j$ for a packet traveling towards $k$ \footnote{It is
assumed that the eigenvalues of the $\mathbf{p}^k$ matrix are smaller
than 1, which must be true if the number of times that a packet goes
through a certain node is finite.}. In particular, the element
$d_{ik}^k$ is the average number of steps needed to find $k$ starting
from $i$. When the search algorithm is such that the packets follow
minimum paths between nodes, the effective distance will coincide with
the topological minimum distance; otherwise, the effective distance
between nodes will be, in general, larger than the topological minimum
distance. Finally, the average search cost in the network when there
is not congestion is
\begin{equation}
\overline{d}=\frac{\sum_{i,k}d_{ik}^k}{S(S-1)},
\label{avd}
\end{equation}
where $S$ is the number of nodes in the network.

Consider next which is the {\it centrality} of each of the nodes in
the communication network. First, we calculate the average number of
times, $b_{ij}^k$, that a packet generated at $i$ and with destination
$k$ passes through $j$. According to the previous definitions
\begin{equation}
\mathbf{b}^k=\sum_{n=1}^\infty \mathbf{P}^k(n)=\sum_{n=1}^\infty
\left(\mathbf{p}^k\right)^n=(\mathbf{I}-\mathbf{p}^k)^{-1}\mathbf{p}^k.
\label{bij}
\end{equation}
The {\it effective} betweenness of node $j$, $B_j$, is defined as
\begin{equation}
B_j=\sum_{i,k}b_{ij}^k.
\label{B}
\end{equation}
Again, as in the case of the effective distance, when the search
algorithm is able to find the minimum paths between nodes, the
effective betweenness will coincide with the {\it topological}
betweenness, $\beta_j$, as usually defined
\cite{freeman77,newman01}. The effective betweenness of the nodes in a
network contains valuable information about its behavior when multiple
searches are performed simultaneously and congestion considerations
become relevant.

Consider the following general scenario.
In the communication network, each node generates packets at a rate
$\rho$ per unit of time, independently of the rest of the nodes.
%
%
The destination of each of these packets is randomly fixed at the
moment of its creation. On the other hand, the nodes are queues that
can store as many packets as needed but can deliver, on average, only
a finite number of them at each time step---without lost of
generality, we fix this number to 1. It is known
\cite{arenas01,ohira98,sole01} that for low values of $\rho$ the
system reaches a steady state in which the total number of {\it
floating} packets in the network $N(t)$ fluctuates around a finite
value. As $\rho$ increases, the system undergoes a continuous phase
transition to a {\it congested phase} in which $N(t)\propto t$
\cite{arenas01}. Right at the critical point, $\rho_c$, quantities
such as $N(t)$ and the characteristic time diverge
\cite{guimera01}. Below $\rho_c$, there is no accumulation at any node
in the network and the number of packets that arrive to node $j$ is,
on average, $\rho B_j/(S-1)$. Therefore, a particular node will
collapse when $\rho B_j/(S-1)>1$ and the critical congestion point of
the network will be
\begin{equation}
\rho_c=\frac{S-1}{B^*}
\label{pc}
\end{equation}
where $B^*$ is the maximum effective betweenness in the network, that
corresponds to the most central node.

To calculate the average of the load of the network, $\langle
N(t)\rangle$, it is necessary to establish the behavior of the
queues. In the general scenario proposed above, the arrival of packets
to a given node $j$ is a Poisson process with mean $\mu_j=\rho
B_j/(S-1)$.
Regarding the delivery of packets, assume the simplest case in which
it is also a Poisson process and hence the time between two
consecutive packet deliveries follows an exponential distribution
\footnote{For example, consider the following model which is a
simplification of the communication model with discrete time proposed
in Ref.~\cite{arenas01}. For a node $j$ with $\nu_j$ packets stored in
its queue, every time step each packet jumps to the next node (chosen
according to the search algorithm defined through the matrices
$\mathbf{p}^k$) with probability $1/\nu_j$. In this simple case, the
delivery of packets can be approximated by a Poisson process. The
continuum approximation holds in this case because the maximum rate of
arrival and delivery is 1 packet per time step and, therefore, the
results obtained in queuing theory can be applied safely. However, the
approximation would not hold for discrete models if larger time
windows were considered.}. In general, when the arrival and delivery
processes are Poisson, the average size of the queues is given by
\cite{allen90}
%
%
\footnote{
Note that, to obtain this result, the order in which packets are
processed is irrelevant.
Moreover, it is straightforward to extend the calculations to other
types of queues.}:

\begin{equation}
\langle\nu_j\rangle=\frac{\mu_j}{1-\mu_j}=\frac{\frac{\rho
B_j}{S-1}}{1-\frac{\rho B_j}{S-1}}.
\end{equation}

The average load of the network $\langle N(t)\rangle$ is
\begin{equation}
\langle N(t)\rangle=\sum_{j=1}^S\langle\nu_j\rangle=\sum_{j=1}^S\frac{\frac{\rho B_j}{S-1}}{1-\frac{\rho B_j}{S-1}}.
\label{load}
\end{equation}

There are two interesting limiting cases of this expression. When
$\rho$ is very small, $\langle\nu_j\rangle\approx\mu_j$ and taking
into account that $\sum_j B_j=\sum_{i,k}d_{ik}^k$, one obtains
\begin{equation}
\langle N(t)\rangle\approx\rho S \overline{d}\quad\quad\rho\rightarrow 0.
\label{N1}
\end{equation}
On the other hand, when $\rho$ approaches $\rho_c$ most of the load of
the network comes from the most congested node, and
therefore\footnote{We assume that there is one only node that has
betweenness $B^*$. If there is more than one, equation (\protect\ref{N2})
should be multiplied by the number of such nodes.}
\begin{equation}
\langle N(t)\rangle\approx\frac{1}{1-\frac{\rho B^*}{S-1}}\quad\quad\rho\rightarrow\rho_c.
\label{N2}
\end{equation}

It is worth noting that there are only two assumptions in the
calculations above. The first one has already been mentioned: the
movement of the packets needs to be Markovian to define the jump
probability matrices $\mathbf{p}^k$. Although this is not strictly true in real
communication networks---where packets are not allowed usually to go
through a given node more than once---it can be seen as a first
approximation \cite{sole01,arenas01,ohira98}. The second assumption is
that the jump probabilities $p_{ij}^k$ do not depend on the congestion
state of the network, although communication protocols sometimes try
to avoid congested regions, and then $B_j=B_j(\rho)$
\footnote{All the derivations above will still be true in a number of
general situations, including situations in which the paths that the
packets follow are unique \cite{arenas01}, in which the routing tables
are fixed, or situations in which the structure of the network is very
homogeneous and thus the congestion of all the nodes is similar}. Our
calculations, in particular equations (\ref{pc})--(\ref{N2}),
correspond to the worst case scenario and thus provide bounds to more
realistic scenarios in which the search algorithm interactively avoids
congestion.

\bigskip

Equations (\ref{bij}), (\ref{B}) and (\ref{load}) enable us to tackle
the problem of finding optimal structures for local search. Optimality
is defined as minimization the average time needed to perform a
search. Indeed, according to Little's Law \cite{allen90}, the average
time needed by a packet to reach its destination is proportional to
the total load of the network, and therefore minimizing $\langle
N(t)\rangle$ is equivalent to minimizing the average cost of a
search. In a local search scenario, the $\mathbf{p}^k$ matrices are
given by
\begin{equation}
p_{ij}^k=a_{ik}\delta_{jk}+(1-a_{ik}-\delta_{ik})\frac{a_{ij}}{\sum_l a_{il}}.
\label{path}
\end{equation}
where $a_{ij}$ are the elements of the adjacency matrix of the
network. The first term corresponds to $i$ and $k$ being neighbors:
then the packet will go to $j$ if and only if $j=k$, i.e. the packet
will be sent directly to the destination. The second term corresponds
to $i$ and $k$ not being neighbors: in this case, $j$ is chosen at
random and uniformly among the neighbors of $i$. Finally, the delta
symbol ensures that $p_{kj}^k=0$ $\forall j$ and the packet {\it
disappears} from the network.

The optimization process is carried out using generalized simulated
annealing as described in \cite{tsallis94,penna95}. Starting from a
given initial network configuration, random rewiring of individual
links are performed, the cost $\langle N(t)\rangle$ is evaluated
according to (\ref{load}) and the change is accepted with a certain
probability that depends on a computational temperature, which is
decreased so that the system tends to explore regions of the
configuration state with lower and lower costs.  Regarding the
cooling, at a given temperature, each node of the network is allowed
to try a rewiring. Then the temperature is decreased by 1\%, and the
process is repeated until a minimum temperature is reached or,
alternatively, the system has remained unchanged after a significantly
large amount of rewiring trials. Different sets of initial conditions
are explored: for a given value of $\rho$, the optimization process is
started from random initial configurations and also from networks that
turned out to be optimal at similar values of $\rho$. Of all the
realizations, only the network with a smallest cost is considered as
optimal.

\begin{figure}[h]
\centerline{\includegraphics*[width=0.5\columnwidth]{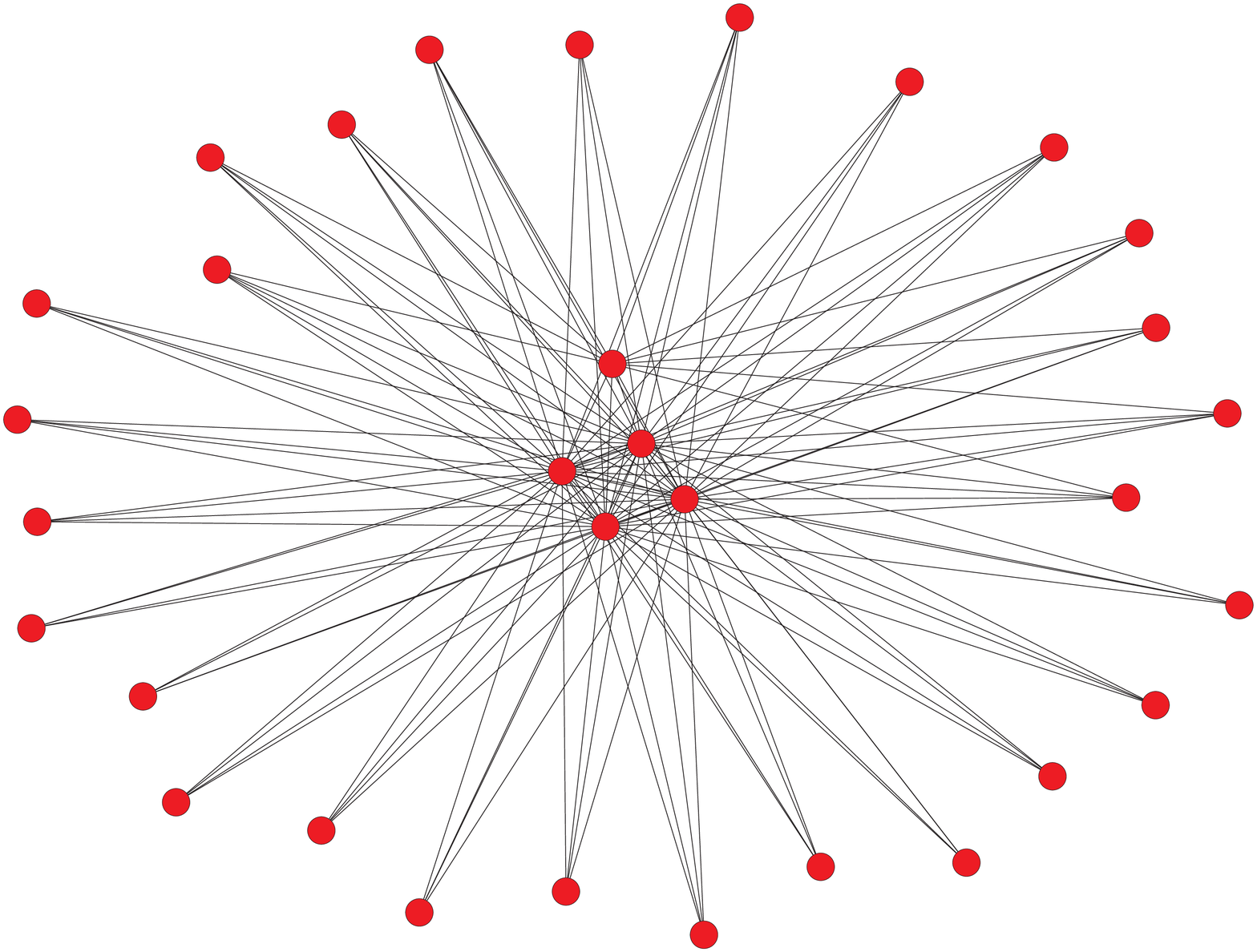}\includegraphics*[width=0.5\columnwidth]{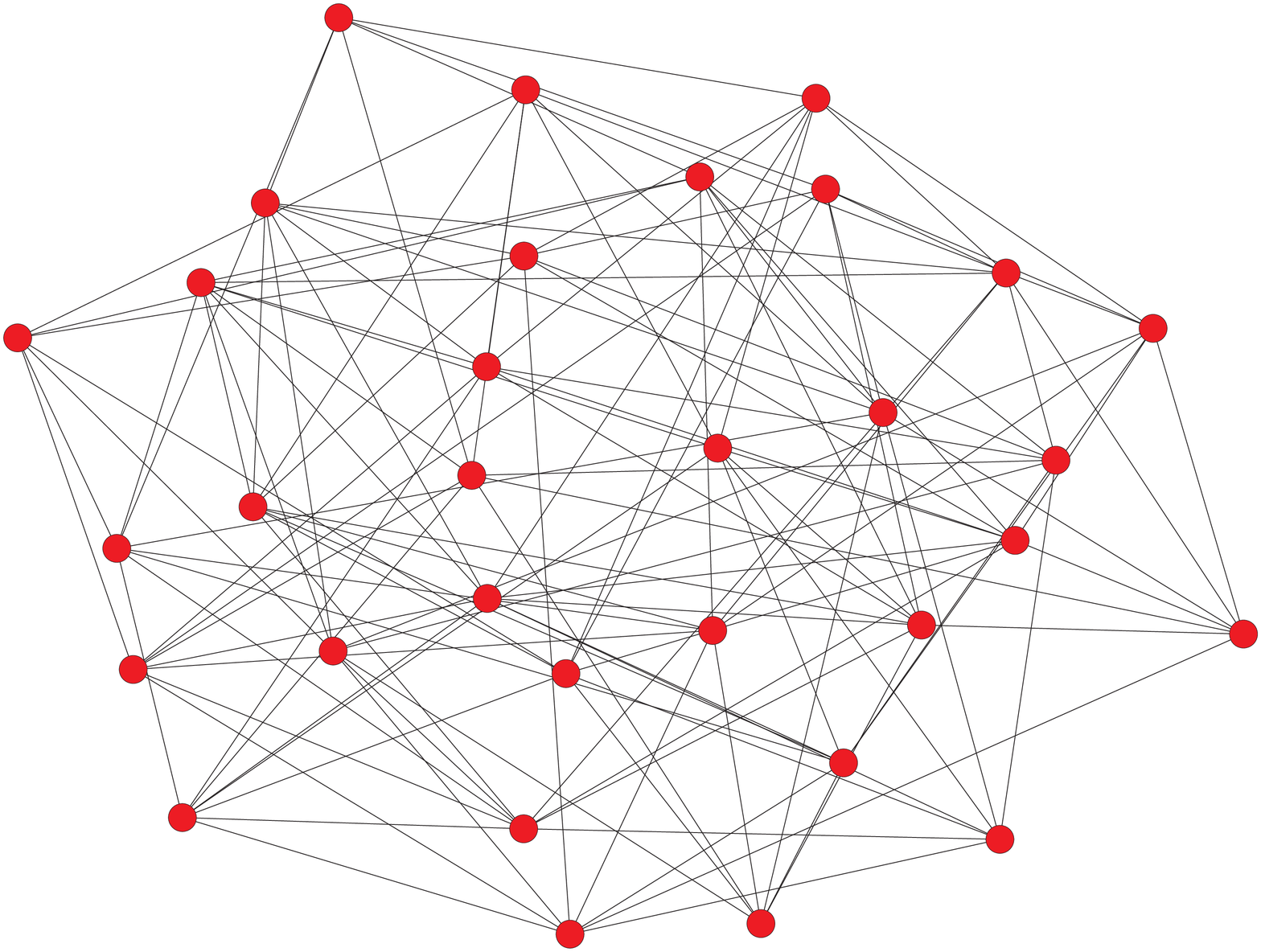}}
\vspace{0.1in}
\centerline{\includegraphics*[width=0.8\columnwidth]{r1}}
\caption{Optimal structures for local search with congestion. Top:
Star-like configuration optimal for $\rho<\rho^*$ (left), and
homogeneous-isotropic configuration optimal for $\rho>\rho^*$
(right). Bottom: Polarization of the optimal structure as a function
of $\rho$, for networks of size $S=32$ and different number of links
$L$.}
\label{optimal}
\end{figure}

The results of the optimization process are shown in Fig.
\ref{optimal}. For $\rho\rightarrow 0$, the optimal network has a
star-like centralized structure as expected, which corresponds to the
minimization of the average effective distance between nodes
(Eq. \ref{N1}). On the other extreme, for high values of $\rho$, the
optimal structure has to minimize the maximum betweenness of the
network, according to Eq. (\ref{N2}). This is accomplished by creating
a homogeneous network where all the nodes have essentially the same
degree, betweenness, etc. To characterize the networks at all values
of $\rho$, we introduce a measure of the {\it polarization}, $\pi$, of
the network:
\begin{equation}
\pi=\frac{\beta^*-\langle\beta\rangle}{\langle\beta\rangle}
\end{equation}
where $\beta$ is, as before, the topological betweenness of the nodes.
For star-like networks, the value of $\pi$ is large while for very
homogeneous networks $\pi\approx 0$. Although one could expect that
optimal networks cover the whole range of values from $\pi=\pi_{star}$
to $\pi\approx 0$, the results of the optimization process reveal a
completely different scenario. According to simulations, star-like
configurations are optimal for $\rho<\rho^*$; at this point, the
homogeneous networks that minimize $B^*$ become optimal. Therefore
there are only two type of structures that can be optimal for a local
search process: star-like networks for $\rho<\rho^*$ and homogeneous
networks for $\rho>\rho^*$.

\bigskip
In summary, we have found analytical expressions for the relationship
between topological properties of networks and the specific dynamic
behavior when faced with local search with congestion. These
expressions allow the calculation of the search cost in terms of the
{\it effective} betweenness of the nodes, which is calculated via the
transition probability matrices (formal expressions of the search
algorithm). This formalism allows to perform an exhaustive search for
optimal topologies in terms of parallel {\it searchability} avoiding
the simulation of the dynamics of the parallel search process, that is
prohibitive in computational time. Moreover, the formalism is general
enough to deal with other search scenarios---local searches with
knowledge up to second nearest neighbors, third nearest neighbors and
so on (eventually, global knowledge)---simply redefining the
$p_{ij}^k$ elements. We find that the optimal network topologies for
local search considering congestion are split in two categories: a
star-like network topology, that is optimal for small number of
parallel searches, and the homogeneous-isotropic network topology,
that is optimal for large numbers of parallel searches. Strikingly,
the transition between these categories is sharp, i.e. we are not able
to find any optimal network topology different from these two classes.


\acknowledgments The authors are grateful to L.~A.~N. Amaral,
L. Danon, X. Guardiola, R. Monasson, C.J. P\'erez, and M. Sales for
helpful comments and discussions. This work has been supported by DGES
of the Spanish Government, Grants No. PPQ2001-1519, No. BFM2000-0626,
No. BEC2000-1029 and No. BEC2001-0980, and EC-Fet Open project
No. IST-2001-33555. R.G. also acknowledges financial support from the
Generalitat de Catalunya.


\end{document}